# Simulation studies on the design of optimum PID controllers to suppress chaotic oscillations in a family of Lorenz-like multi-wing attractors

Saptarshi Das[a,b], Anish Acharya[c,d], and Indranil Pan[a]

a) Department of Power Engineering, Jadavpur University, Salt Lake Campus, LB-8, Sector 3, Kolkata-700098, India.
b) Communications, Signal Processing and Control (CSPC) Group, School of Electronics and Computer Science, University of Southampton, Southampton SO17 1BJ, United Kingdom.
c) Department of Instrumentation and Electronics Engineering, Jadavpur University, Salt Lake Campus, LB-8, Sector 3, Kolkata-700098, India.
d) Department of Electrical and Computer Science Engineering, University of California Irvine, CA 92697-2620, USA.

**Authors' Emails:**
saptarshi@pe.jusl.ac.in, s.das@soton.ac.uk (S. Das)
anisha@uci.edu (A. Acharya)
indranil.jj@student.iitd.ac.in, indranil@pe.jusl.ac.in (I. Pan)

**Abstract:**
Multi-wing chaotic attractors are highly complex nonlinear dynamical systems with higher number of index-2 equilibrium points. Due to the presence of several equilibrium points, randomness and hence the complexity of the state time series for these multi-wing chaotic systems is much higher than that of the conventional double-wing chaotic attractors. A real-coded Genetic Algorithm (GA) based global optimization framework has been adopted in this paper as a common template for designing optimum Proportional-Integral-Derivative (PID) controllers in order to control the state trajectories of four different multi-wing chaotic systems among the Lorenz family *viz.* Lu system, Chen system, Rucklidge (or Shimizu Morioka) system and Sprott-1 system. Robustness of the control scheme for different initial conditions of the multi-wing chaotic systems has also been shown.

## 1. Introduction

Chaos is a field in mathematical physics which has found wide applications, ranging from weather prediction to geology, mathematics, biology, computer science, economics, philosophy, sociology, population dynamics, psychology, robotics etc. [29]. Chaos theory studies the behavior of dynamical systems which are nonlinear, highly sensitive to changes in initial conditions, and have deterministic (rather than probabilistic) underlying rules which dictates the evolution of the future states of a system. Such systems exhibit aperiodic oscillations in the time series of the state variables and long term prediction of such systems is not possible due to the sensitive dependence on initial conditions. Theoretical investigations of chaotic time series have percolated into real world applications like those in computer networks, data encryption, information processing systems, pattern recognition, economic forecasting, stock market prediction etc. [29]. Historically, chaos was first observed in natural convection [16], weather prediction and subsequently in several other fluid mechanics related problems [2]. Survey of a wide variety of physical systems exhibiting chaos and pattern formation in nature ranging from convection, mixing of fluids, chemical reaction, electro-hydrodynamics to solidification and nonlinear optics have been summarized in [10].

An interesting ramification of chaos theory is the investigation of synchronization and control of such systems [19]. Even with their unpredictable dynamics, proper control laws can be designed to suppress the undesirable excursions of the state variables or make the state variables follow a definite trajectory [35]. These chaos control schemes have a variety of engineering applications and hence control of chaotic systems has received increased attention in last few years [5]. In chaos control, the prime objective is to suppress the chaotic oscillations completely or reduce them to regular oscillations. Recent chaos control techniques include open loop control methods, adaptive control methods, traditional linear and nonlinear control methods, fuzzy control techniques etc. among many others [5]. Most chaos control techniques exploit the fact that any chaotic attractor containing infinite number of unstable periodic orbits can be modified using an external control action to produce a stable periodic orbit. The chaotic system's states never remains in any of these unstable orbits for a long time but rather it continuously switches from one orbit to the other which gives rise to this unpredictable, random wandering of the state variables over a longer period of time. The control scheme basically tries to achieve stabilization, by means of small system perturbations, of one of these unstable periodic orbits. The result is to render an otherwise chaotic motion more stable and predictable. The perturbation must be tiny, to avoid significant modification of the system's natural dynamics. Several techniques have been used to control chaos, but most of them are developments of two basic approaches: the OGY (Ott, Grebogi and Yorke) method [20], and Pyragas continuous control method [22]. Both methods require a prior determination of the unstable periodic orbits of the chaotic system before the control algorithm can be designed. The basic difference between the OGY and Pyragas methods of chaos control is that the former relies on the linearization of the Poincare map and the latter is based on time delay feedback.

PID controller based designs are popular in the control engineering community for several decades, due to its design simplicity, ease of use, and implementation feasibility. PID type controller design can be found in recent literatures for synchronization of chaotic systems with different initial guess [1, 3, 7-9, 14, 15]. However, except few attempts like [4], optimum PID controller design for chaos control is not well investigated yet, especially for the control of highly complex chaotic systems like multi-wing attractors, as attempted in this paper. The rationale behind particularly choosing PID controllers is due to its simplicity, ease of implementation and tuning methods compared to unavailability of Lypunov based stabilization and sliding mode techniques [19, 35] for multi-wing chaotic attractors. The classical approach of dealing with chaos control and synchronization is to formally eliminate the nonlinear terms for designing a suitable nonlinear control scheme, in order to make the error dynamical system linear and then find a suitable Lypunov based stabilization scheme [19]. The present approach needs no additional requirement of canceling complex multi-segment nonlinear terms by a nonlinear control scheme for each chaotic system. The success of the present approach lies in the fact that a generalized linear PID control framework can stabilize a family of complex multi-wing chaotic system when only the second state variable needs to be sensed and manipulated using an external PID control action. The controller gains are optimized to damp the oscillations as early as possible and the control scheme is also tested for required robustness against system's initial condition variation. Previously, design of PID type controllers equipped with fractional order integro-differential operators and fuzzy logic have been researched for chaos synchronization [11] and chaos control [21], within a common global optimization framework without paying much attention to the specific nonlinear structure and complexity of the chaotic system. This idea has been extended here for the control of highly complex nonlinear dynamical systems i.e. multi-wing chaotic attractors in the Lorenz family.

Many researches have focused on theoretical and experimental studies on the synchronization and control of classical double-wing chaotic and hyper-chaotic systems from the Lorenz family like the Lorenz, Chen and Lu attractors [19, 35, 5, 12]. However, there is almost no significant result for relatively complex chaotic systems like the multi-scroll and multi-wing attractors which is addressed in the present paper. Multi-scroll attractors first emerged as an extension of the Chua's circuit [31] and it has been found to be more suitable than the classical double-wing chaotic attractors in applications like secure communication and data encryption. The multi-scroll attractors are produced by suitably modifying the nonlinear terms of the governing differential equations of the chaotic system which has been successfully applied to produce 1D, 2D and 3D scroll grid attractors in [17]. The reason behind obtaining highly complex time-series out of multi-scroll and multi-wing attractors is that they have more number of equilibrium points than the classical double-wing attractors. Recent research has shown that such highly complex time-series and phase space behavior can also be obtained from typical chaotic and hyper-chaotic systems with no equilibrium point [27] and infinite number of equilibrium points [36]. Theoretical studies on the complexity analysis of such multi-wing attractors have been reported in He *et al.* [13] using spectral entropy and statistical complexity measure. Among the two major families of complex chaotic systems, the control and synchronization of multi-scroll attractors are proposed in [30] but there is almost no result for the multi-wing attractors.

The purpose of the present study is to develop a common template to suppress highly complex chaotic oscillations in the Lorenz family of multi-wing attractors using a simple controller structure like the PID and a global optimization based gain tuning mechanism. An objective function for the controller design problem is framed in terms of the state variables of the chaotic system and the desired state trajectories of the system. The real coded GA based optimization is then employed to find out the values of the PID controller gains.

## 2. Basics of the Lorenz family of multi-wing chaotic systems

Four classical examples of symmetric double-wing chaotic attractors are chosen as the base cases here among the Lorenz family to study the control of multi-wing attractors [32]. In spite of having several literatures on the generation of multi-scroll attractors as an extension of Chua's circuit, the first extension of the Lorenz family of attractors from double-wing to multi-wing was proposed by Yu *et al.* [32]. It is also observed that in the Lorenz family of chaotic systems, similar attractors could be generated by replacing the cross-products with quadratic terms. The common characteristics of the source of nonlinearity in the state equations of the original double-wing Lorenz family of chaotic systems are due to having either a square and/or cross-terms of the state variables. These particular terms can be replaced by a multi-segment parameter adjustable quadratic function (1) to generate multi-wing attractor with additional flexibility of modifying the numbers and locations of index-2 equilibrium points. As reported in the pioneering works of multi-wing chaos [32]-[33], the segment characteristics like the slope and width can be adjusted using the parameters $\{F_0, F_i, E_i\}$ of equation (1). This typical function increases the number of index-2 equilibrium points along a particular axis for Lorenz family of chaotic systems from two to $(2N+2)$, thereby increasing the randomness or complexity of the state trajectories of the nominal (double-wing) chaotic system which is difficult to control by analytical methods.

$$f(x) = F_0 x^2 - \sum_{i=1}^{N} F_i \left[ 1 + 0.5 \operatorname{sgn}(x - E_i) - 0.5 \operatorname{sgn}(x + E_i) \right] \tag{1}$$

where,

$$\operatorname{sgn}(x) = \begin{cases} 1 & \text{for } x > 0 \\ 0 & \text{for } x = 0 \\ -1 & \text{for } x < 0 \end{cases} \tag{2}$$

and $N$ being a positive integer responsible for the number of wings of the chaotic attractor.

In this paper, two multi-wing chaotic systems among the Lorenz family are obtained by replacing the cross-product terms (in Lu and Chen system) by the above mentioned multi-segment function. In the other two systems among the Lorenz family, the quadratic terms (in Rucklidge and Sprott-1 system) are replaced by the multi-segment function (1).

### *2.1. Chaotic multi-wing Lu system*

The double-wing chaotic Lu system [18] is represented by (3).

$$\begin{aligned} \dot{x} &= -ax + ay \\ \dot{y} &= cy - xz \\ \dot{z} &= xy - bz \end{aligned} \tag{3}$$

The typical parameter settings for chaotic double-wing Lu attractor is given as $a = 36, b = 3, c = 20$. The equilibrium points of the Lu system are located at $(0,0,0); (\pm\sqrt{bc}, \pm\sqrt{bc}, c)$. The state equations of the multi-wing chaotic Lu attractor whose states are to be controlled using a PID controller are given by (4).

$$\begin{aligned} \dot{x} &= -ax + ay \\ \dot{y} &= cy - (1/P)xz + u \\ \dot{z} &= f(x) - bz \end{aligned} \tag{4}$$

Here, $P$ reduces the dynamic range of the attractors so as to facilitate hardware realization. The suggested parameters for $N = 4$ are given in (5) [32] for which the chaotic Lu system exhibits multi-wing attractors in the phase portraits.

$$P = 0.05, F_0 = 100, F_1 = 10, F_2 = 12, F_3 = 16.67, F_4 = 18.18,$$
$$E_1 = 0.3, E_2 = 0.45, E_3 = 0.6, E_4 = 0.75 \tag{5}$$

Here, the original cross product ($xy$) is replaced by the multi-segment quadratic function $f(x)$. Fig. 1 shows the projections of the 3-dimensional (*x-y-z*) phase space dynamics (Fig. 1d) of the multi-wing Lu system on the *x-y* (Fig. 1a), *y-z* (Fig. 1b), *x-z* (Fig. 1c) planes. Also, in (4) the PID control action is added to the second state variable to suppress the chaotic oscillations and the control action is given by (6).

$$u = K_p e + K_i \int e.dt + K_d \frac{de}{dt}, \quad e = (r - y) \tag{6}$$

Here, $\{K_p, K_i, K_d\}$ are the PID controller gains which are to be found out by a suitable optimization technique for the reference signal ($r$) as a unit step.

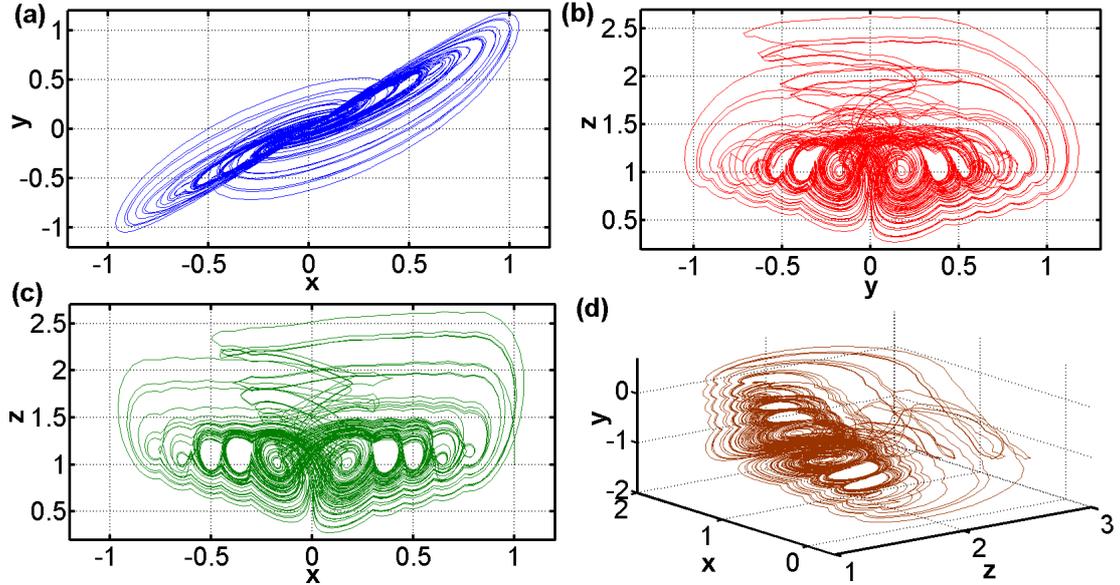

Fig. 1. Uncontrolled phase plane portraits for multi-wing Lu system [32].

### 2.2. Chaotic multi-wing Chen system

The double-wing Chen system [6] is represented by (7).

$$\dot{x} = -ax + ay$$
$$\dot{y} = (c - a)x + cy - xz \tag{7}$$
$$\dot{z} = xy - bz$$

The typical parameter settings for chaotic double-wing Chen attractor is given as $a = 35, b = 3, c = 28$. The equilibrium points of the Chen system are located at $(0,0,0); \left(\pm\sqrt{b(2c-a)}, \pm\sqrt{b(2c-a)}, (2c-a)\right)$. The state equations of the multi-wing chaotic Chen attractor whose states are to be controlled are given by (8).

$$\dot{x} = -ax + ay$$
$$\dot{y} = (c-a)x + cy - (1/P)xz + u \quad (8)$$
$$\dot{z} = f(x) - bz$$

The suggested parameters for $N = 4$ are given in (9) for which the chaotic Chen system exhibits multi-wing attractors in phase portraits.

$$P = 0.05, F_0 = 100, F_1 = 12, F_2 = 12, F_3 = 16.67, F_4 = 18.75,$$
$$E_1 = 0.3, E_2 = 0.45, E_3 = 0.6, E_4 = 0.75 \quad (9)$$

The multi-wing Chen system in Fig. 2 shows similar phase space behavior to that of the multi-wing Lu system in Fig. 1, except the fact that more number of trajectories are observed away from the equilibrium points. Similar to the previous case of multi-wing Lu system, the nonlinear cross-product ($xy$) in the third state variable is replaced by $f(x)$.

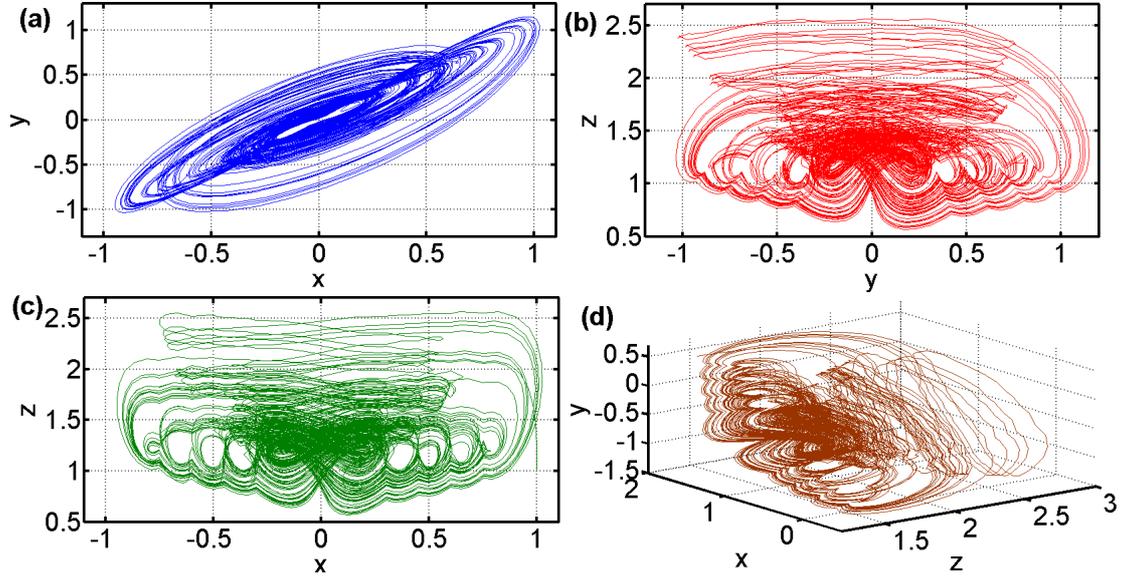

Fig. 2. Uncontrolled phase plane portraits for multi-wing Chen system [32].

## 2.3. Chaotic Multi-wing Rucklidge system

The double-wing Shimizu-Morioka system [24] is given by (10).
$$\dot{x} = -ax + by - yz$$
$$\dot{y} = x \quad (10)$$
$$\dot{z} = y^2 - z$$

The typical parameter settings for chaotic double-wing Shimizu-Morioka attractor is given as $a = 2, b = 7.7$. The equilibrium points of the Shimizu-Morioka system are located at $(0,0,0); (0, \pm\sqrt{b}, b)$. The state equations of the multi-wing chaotic Shimizu-Morioka attractor whose states are to be controlled are given by (11).

$$\dot{x} = -ax + by - (1/P)yz$$
$$\dot{y} = x + u \quad (11)$$
$$\dot{z} = f(y) - z$$

The above mentioned multi-wing chaotic Shimizu-Morioka system [34] is also known as modified Rucklidge system [23]. The suggested parameters for $N = 3$ are given in (12) [32] for which the chaotic Rucklidge system exhibits multi-wing attractors in the phase portraits (Fig. 3).

$$P = 0.5, F_0 = 4, F_1 = 9.23, F_2 = 12, F_3 = 18.18, E_1 = 1.5, E_2 = 2.25, E_3 = 3.0 \quad (12)$$

Here in the case of Rucklidge system, the quadratic term ($y^2$) (unlike the cross-product of states in previous cases of Lu and Chen system) is replaced by the multi-segment function $f(y)$ in the third state equation.

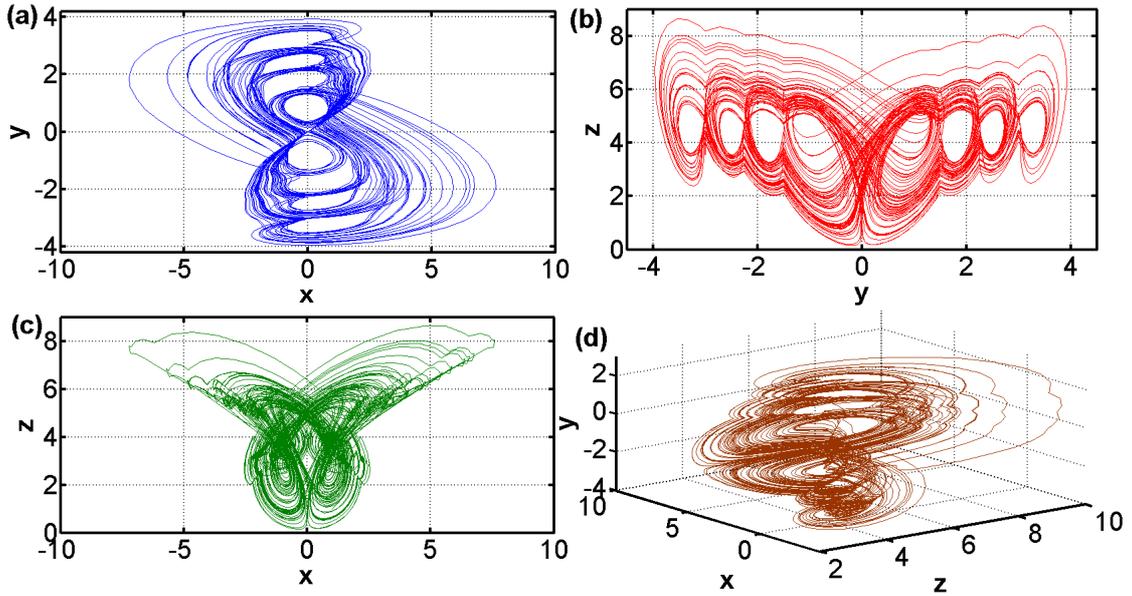

Fig. 3. Uncontrolled phase plane portraits for multi-wing Rucklidge (Shimizu-Morioka) system [32].

### 2.4. Chaotic Multi-wing Sprott-1 system

The double-wing Sprott-1 system [25]-[26] is given by (13).

$$\begin{aligned}
\dot{x} &= yz \\
\dot{y} &= x - y \\
\dot{z} &= 1 - x^2
\end{aligned} \quad (13)$$

The equilibrium points of the Sprott-1 system are located at $(\pm 1, \pm 1, 0)$. State equations of the multi-wing chaotic Sprott-1 attractor whose states are to be controlled are given by (14).

$$\begin{aligned}
\dot{x} &= yz \\
\dot{y} &= x - y + u \\
\dot{z} &= 1 - f(x)
\end{aligned} \quad (14)$$

The suggested parameters for $N = 4$ are given by (15) [32] for which the chaotic Sprott-1 system exhibits multi-wing attractors in phase portraits (Fig. 4).

$$F_0 = 1, F_1 = 5, F_2 = 5, F_3 = 6.67, F_4 = 8.89, E_1 = 2, E_2 = 3, E_3 = 4, E_4 = 5 \quad (15)$$

Here, the quadratic term ($x^2$) is replaced by the multi-segment function $f(x)$ similar to the previous case of Rucklidge system. It is evident from the phase portraits of all the four multi-wing chaotic systems that the wandering of the states in phase space are highly complex, thus indicating the corresponding state time series being highly jittery which is hard to regularize using analytically derived external control action. Here, we have shown that a simple PID type linear controller structure which is widely used in industrial process control is capable of suppressing chaotic oscillations in such complex dynamical systems.

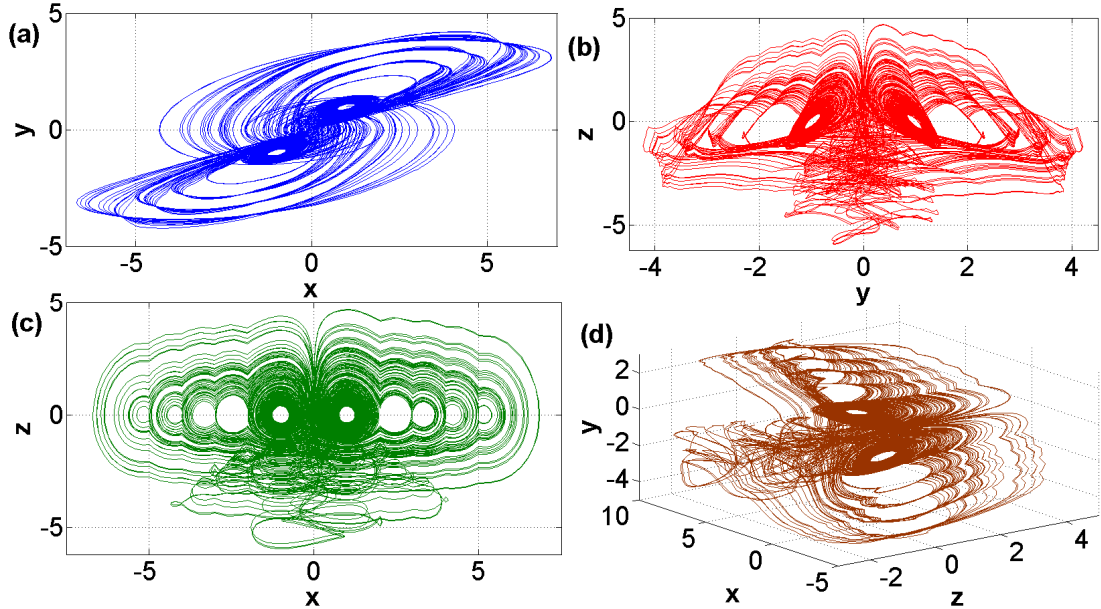

Fig. 4. Uncontrolled phase plane portraits for multi-wing Sprott-1 system [32].

## 3. Optimum PID control to suppress chaotic oscillations in multi-wing attractors

Each of the above four classes of multi-wing chaotic systems are to be controlled using a PID controller (6) which will enforce the second state variable ($y$) to track a unit reference step input ($r$). Instead of simple error minimization criteria for PID controller tuning, the well-known Integral of Time multiplied Absolute Error (ITAE) criterion has been used as the performance index $J$ (16) in order to ensure fast tracking of the second state variable with lesser oscillations.

$$J = \int_0^T t \cdot |e(t)| dt, \quad e(t) = r(t) - y(t) \tag{16}$$

For time domain simulations, the upper time limit of the above integral ($T$) is restricted to realistic values depending on the speed of the chaotic time series to ensure that all oscillations in the state variables have died down due to introduction of the PID control action in the second state. It has also been shown through simulation examples that controlling the second state variable with PID automatically damps chaotic oscillations in the other two state variables for the Lorenz family of multi-wing attractors.

Tuning of the PID controller gains have been done in this study using the widely used population based global optimizer known as the real coded genetic algorithm [28]. The GA is a stochastic optimization process which can be used to minimize a chosen objective function. A solution vector is initially randomly chosen from the search space and undergoes reproduction, crossover and mutation, in each generation, to give rise to a better population of solution vectors in the next generation. Reproduction implies that solution vectors with higher fitness values can produce more copies of themselves in the next generation. Crossover refers to information exchange based on probabilistic decisions between solution vectors. In mutation a small randomly selected part of a solution vector is occasionally altered, with a very small probability. This way the solution is refined iteratively until the objective function is minimized below a certain tolerance level or the maximum number of generations are exceeded. In this study, the population size in GA is chosen to be 20. The crossover and mutation fraction are chosen to be 0.8 and 0.2 respectively for the present minimization problem using (16).

Due to the randomness of the chaotic time series of the multi-wing attractors, the error signal with respect to step command input also becomes highly jittery and will contain several minima which justifies the application of GA in such controller tuning problems as compared to other gradient based optimization algorithms. GA being a global optimization algorithm is able to get out of the local minima, whereas the other gradient based optimization algorithms often get stuck in the local minima and cannot give good solutions. For the control and synchronization of chaotic systems, an evolutionary and swarm based PID controller design with other time domain performance index optimization have been used previously as reported in [11], [21]. But for the sake of simplicity, we have restricted the study for ITAE based PID design only to handle multi-wing attractors in chaotic nonlinear dynamical systems. The GA based optimization results for the PID controller parameters (gains) are given in Table 1 for the four respective multi-wing attractors among the Lorenz family. Also, in order to ensure that the best possible solution is found in the global optimization process, the algorithm has been run several times and only the controller gains, resulting in the fastest tracking performance (and hence the lowest value of $J_{min}$) for respective systems are reported here.

Table 1: GA Based Optimum PID Controller Settings for Chaos Suppression in Multi-wing Attractors

| **Multi-wing Chaotic systems** | $J_{min}$ | $K_p$ | $K_i$ | $K_d$ |
|---|---|---|---|---|
| Lu system | 244.986 | 3.156 | 27.562 | 1.449 |
| Chen system | 307.709 | 3.305 | 21.274 | 1.591 |
| Rucklidge system | 1.161 | 19.435 | 30.097 | 0.237 |
| Sprott-1 system | 1468.193 | 0.272 | 0.433 | 0.393 |

In all cases except the Sprott-1 system, the integral gains ($K_i$) take high value signifying that the controller gives extra effort to reduce the steady state error using the integral action. Also, due the sluggish nature of the uncontrolled dynamics of Rucklidge system, the proportional gain ($K_p$) also takes high value to make the overall closed loop system faster. It is obvious that such controller gains justifies the final objective of ITAE (16) that puts extra penalty for sluggish controlled response and pushes the system to reduce oscillations as soon as possible. A simpler Integral of Absolute Error (IAE) error criterion without the temporal effect in the objective function, unlike ITAE, would have yielded different controller parameters since it only enforces tracking of the set-point but not the fastest possible set-point tracking.

## 3.1. PID Control of Multi-wing Lu System

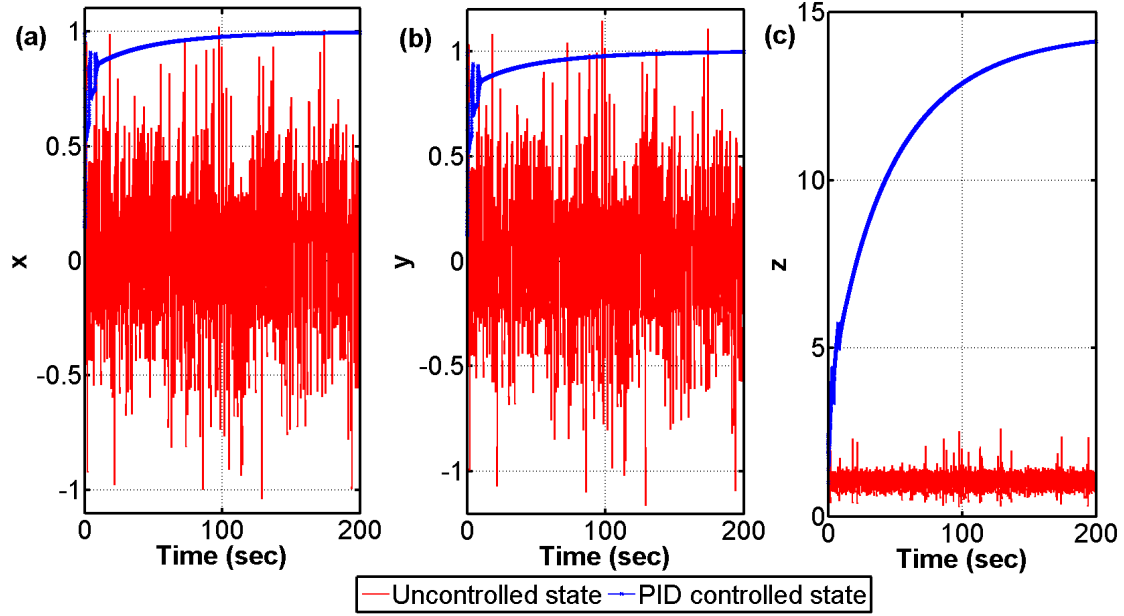

Fig. 5. Uncontrolled and PID controlled response of the state variable for multi-wing Lu system (a) state-*x* (b) state-*y* (c) state-*z*.

Simulation results for the uncontrolled and PID controlled state variables of the multi-wing Lu system (4) has been shown in Fig. 5 with the corresponding control signal and error of the second state depicted in Fig. 6. In Fig. 5(a)-(c), it can be observed that the chaotic evolution of the state variables in the uncontrolled state disappears when the PID controller is applied to the system. All the controlled states evolve to a steady state value. It is evident that the second state variable of the multi-wing Lu system not only tracks the unit step reference but the randomness in the other two state variables is also stabilized. As can be observed from Fig. 5, the controller achieves its design objective of suppressing the oscillations in the second state and making it track the desired trajectory of a unit step. Although the oscillations in the other states x and z are suppressed, the final value at which they settle is dependent on the controller parameters and cannot be obtained a-priori. A more customized control action could have been incorporated like that in [11] utilizing the sum of error for all the three state variables with the intention of getting set-point tracking in all the three state variables. But it would unnecessarily increase the control action which will create difficulty in implementing the control scheme in an analog circuit. A closer look on the set-point tracking error in Fig. 6 shows that the error asymptotically goes to zero and correspondingly the PID control action stabilizes to its final value. This essentially implies that in the steady state, the controller has to continuously give a constant signal to the second state of the chaotic system so that the error remains zero and the oscillations are suppressed.

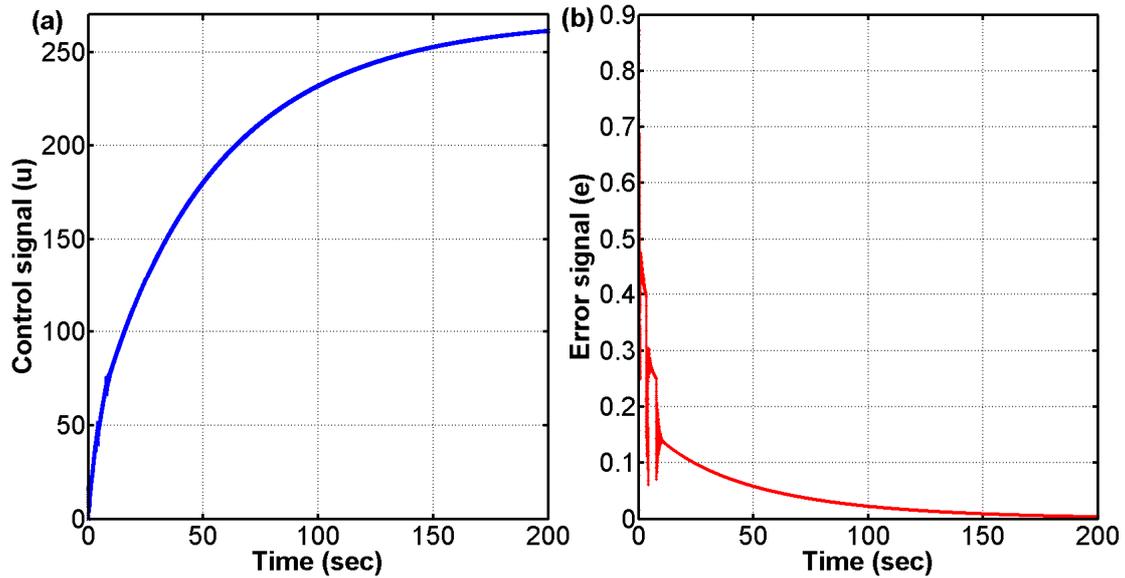

Fig. 6. Control signal and the second state error for the PID controlled multi-wing Lu system.

Also, the time series of the controlled states in Fig. 5(a)-(c) show initial oscillations which get damped quite fast even with a simple GA based optimum PID controller having a linear structure, though better performance can be expected at the cost of implementation of complex nonlinear controller structures and associated computational complexity like that used in [21].

### *3.2. PID control of multi-wing Chen system*

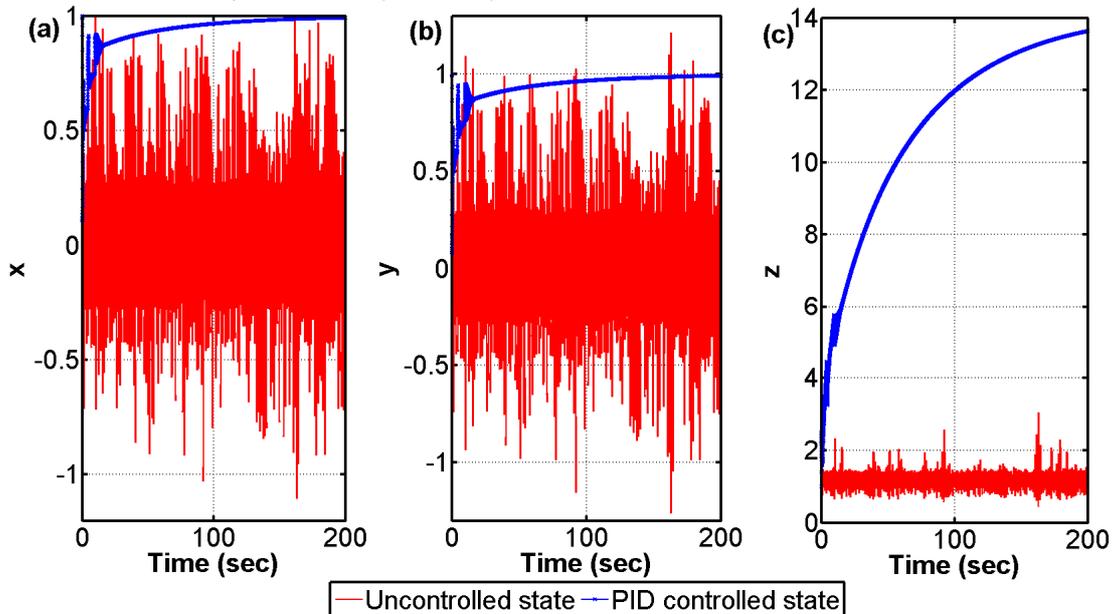

Fig. 8. Uncontrolled and PID controlled response of the state variables for multi-wing Chen system (a) state-$x$ (b) state-$y$ (c) state-$z$.

Fig. 8 shows the controlled three state trajectories of the multi-wing Chen system (8). The associated error signal and control signal (Fig. 9) are quite similar to that of the Lu system. Damping of the chaotic oscillations in the PID controlled phase space is shown in Fig. 10 for this particular system among the Lorenz family. Similar to the case of Lu system, control of Chen system also shows initial jerk in the state trajectories which finally settles down asymptotically. A comparison can be made between the control efforts of Fig. 6 and Fig. 9. It can be observed that the area under the curve is less for Fig. 9 (a) than Fig. 6 (a). This implies that the overall control effort required for controlling the multi-wing Chen system is less than that required for controlling the multi-wing Lu system.

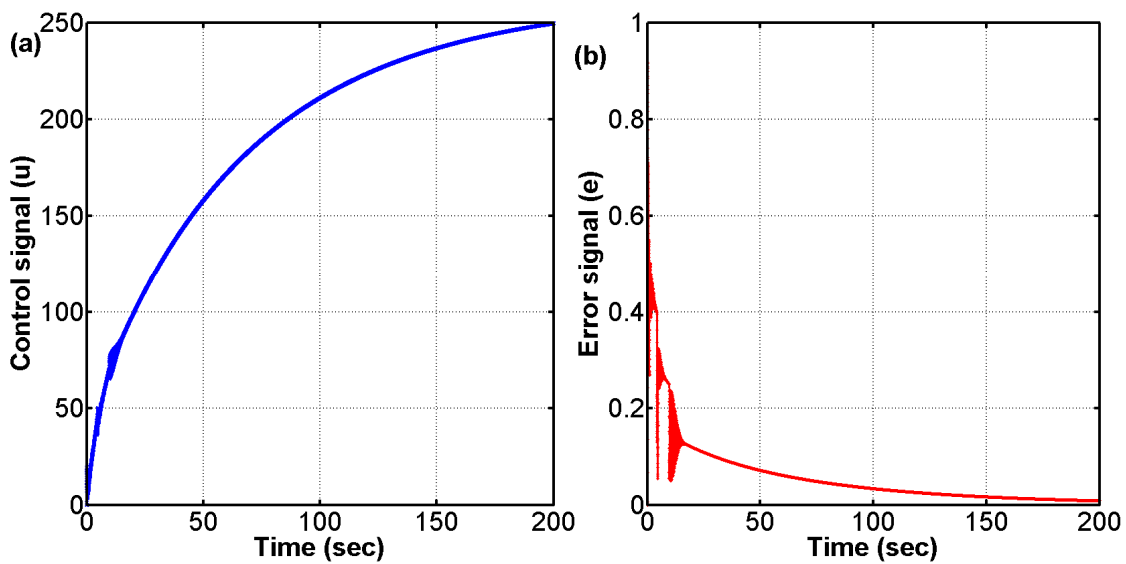

Fig. 9. Control signal and the second state error for the PID controlled multi-wing Chen system.

### 3.3. PID control of multi-wing Rucklidge system

Similar nature of chaos control can be obtained in the multi-wing Rucklidge system (11) also with the GA based optimum PID controller which enforces fast tracking of the second state variable. Also the irregular oscillations of this system are found to be more sluggish compared to the Lu system which is controlled by the PID to track a reference input using ITAE criterion. Also, controlled state trajectories are smoother at initial stages unlike that for the Lu and Chen system. Here, Fig. 11 shows the respective uncontrolled and controlled three state trajectories. The control and error signals are shown in Fig. 12. A comparison can be made between Fig. 11 and Figs. 5 and 8. It can be seen that the oscillations in the multi-wing Rucklidge system is suppressed within a few seconds of applying the control signal, unlike that in the two former cases. A comparison of the control and error signals in this case (Fig. 12) as compared to the previous cases (Figs. 6 and 9) show that the steady state control signal required in this case is much smaller than the former two. The error goes to zero within the first 10 seconds while it takes almost 200 seconds in the former two cases. Therefore it can be concluded that the Rucklidge system is comparatively easier to control using a simple linear PID controller structure than the previously studied chaotic systems i.e. Lu and Chen system.

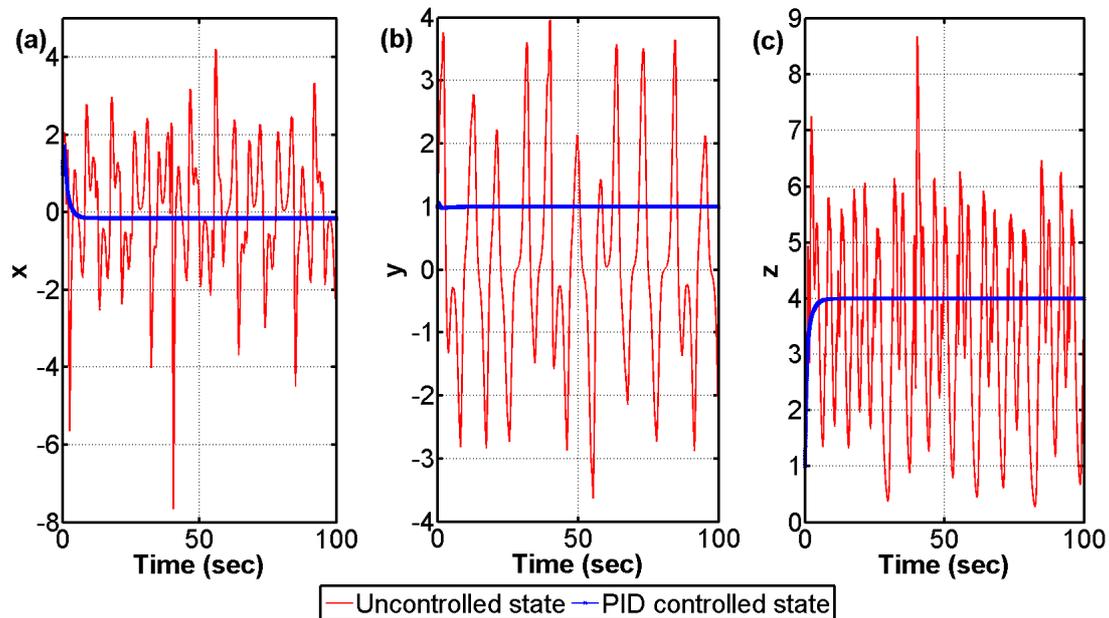

Fig. 11. Uncontrolled and PID controlled response of the state variables for multi-wing Rucklidge system (a) state-*x* (b) state-*y* (c) state-*z*.

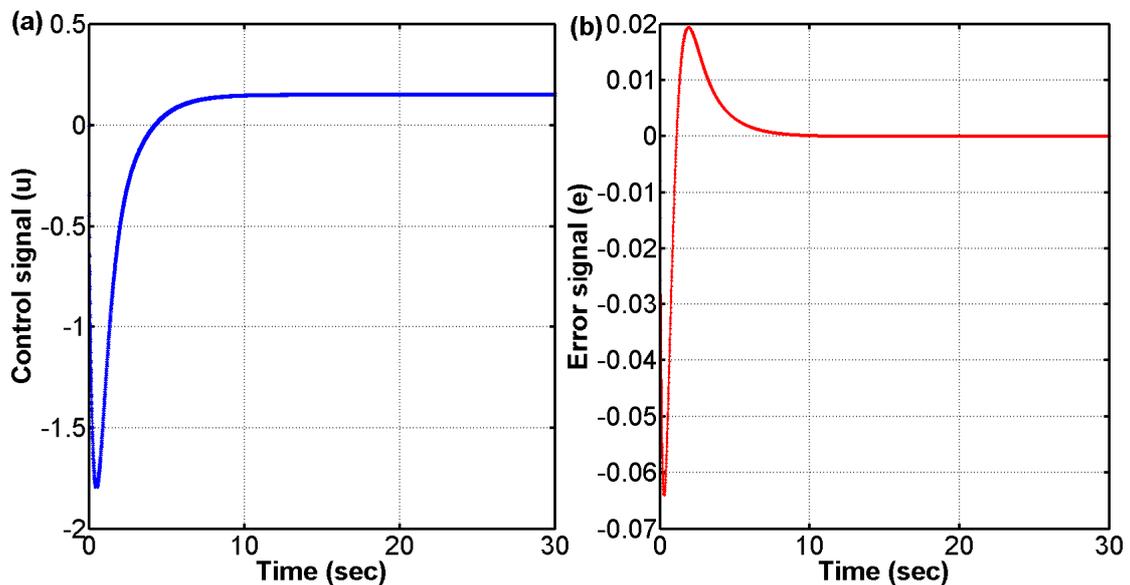

Fig. 12. Control signal and the second state error for the PID controlled multi-wing Rucklidge system.

### *3.4. PID control of multi-wing Sprott-1 system*

For the multi-wing Sprott-1 system (14), the ITAE based GA tuned PID enforces fast reference tracking and simultaneously damps chaotic oscillations (Fig. 14) in an efficient way. It can be noted that in this case, although the original objective of making the second state (*y*) follow a step function is achieved, there are small oscillations in the *x* and the *z* states. This is unlike the previous three cases where controlling one of the states resulted in controlling all the other states automatically. If the oscillations in the other two states are desired to be minimized then they have to be explicitly taken into account in the objective function itself.

It is also evident from the control and error signals in Fig. 15 (which shows the settling of the error signal to zero and control action approaching to its final value) that this system is easier to control than the multi-wing Chen and Lu system. But it is more difficult to control than the multi-wing Rucklidge system, in the sense that it takes more control effort than the former and also takes much more time to settle than the former.

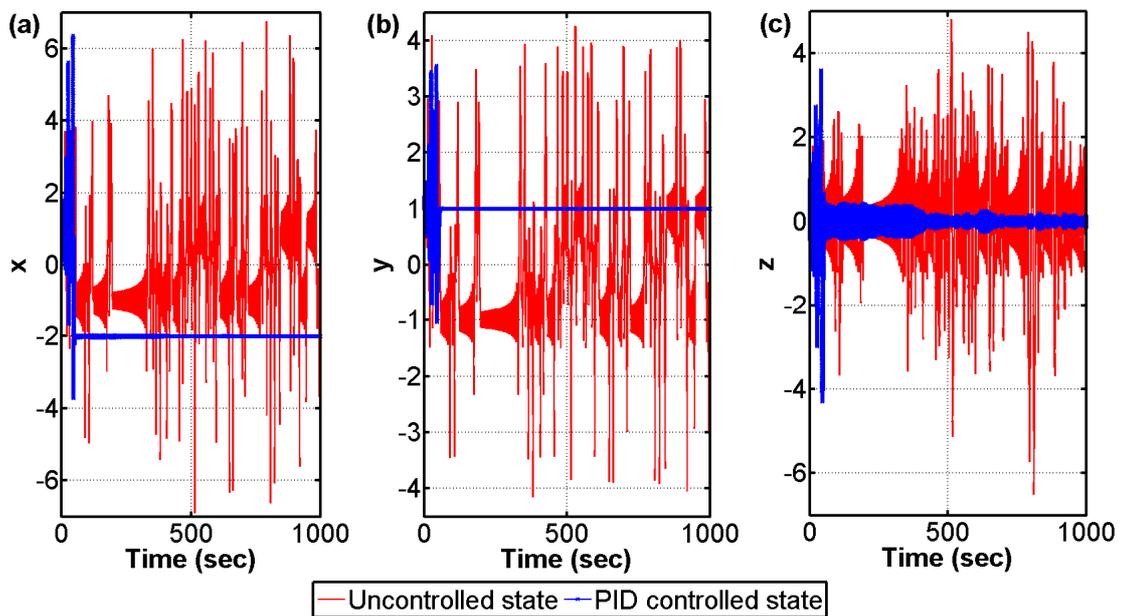

Fig. 14. Uncontrolled and PID controlled response of the state variables for multi-wing Sprott-1 system (a) state-*x* (b) state-*y* (c) state-*z*.

It is well known that chaotic systems are highly sensitive to the initial conditions of the states. Since in the presented approach only a single initial condition of the state variables are assumed to tuned the GA based PID controllers with minimum ITAE for the second state, hence under different initial conditions, effective damping of chaotic oscillations need to be investigated. Therefore, study of the robustness of the present PID control scheme for effective control is shown in next section for each of the four classes of multi-wing chaotic systems.

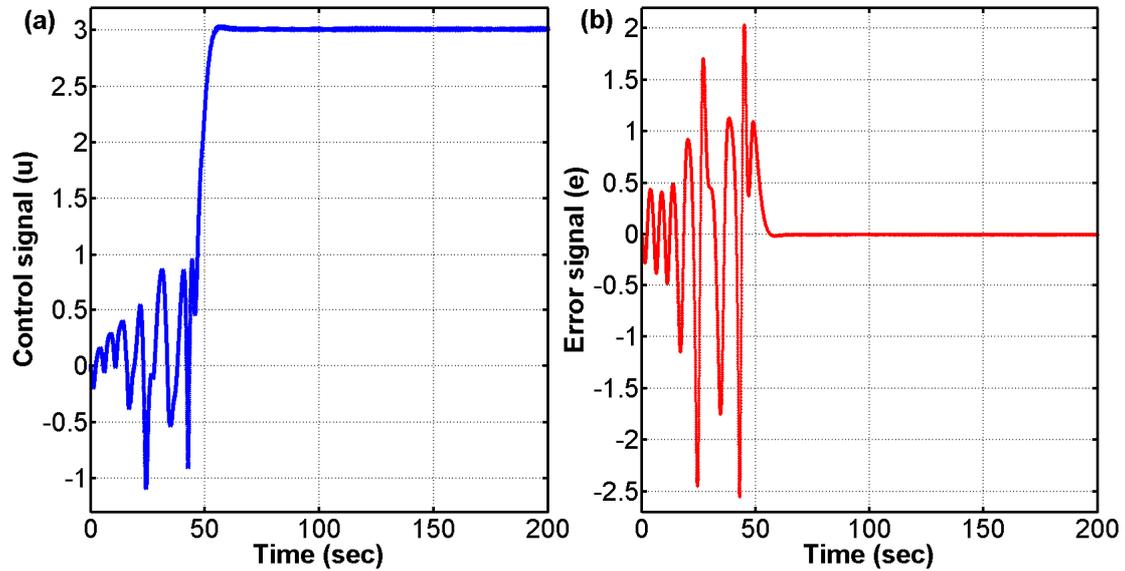

Fig. 15. Control signal and the second state error for the PID controlled multi-wing Sprott-1 system.

## 4. Test of robustness for the PID control scheme with different initial conditions of state variables

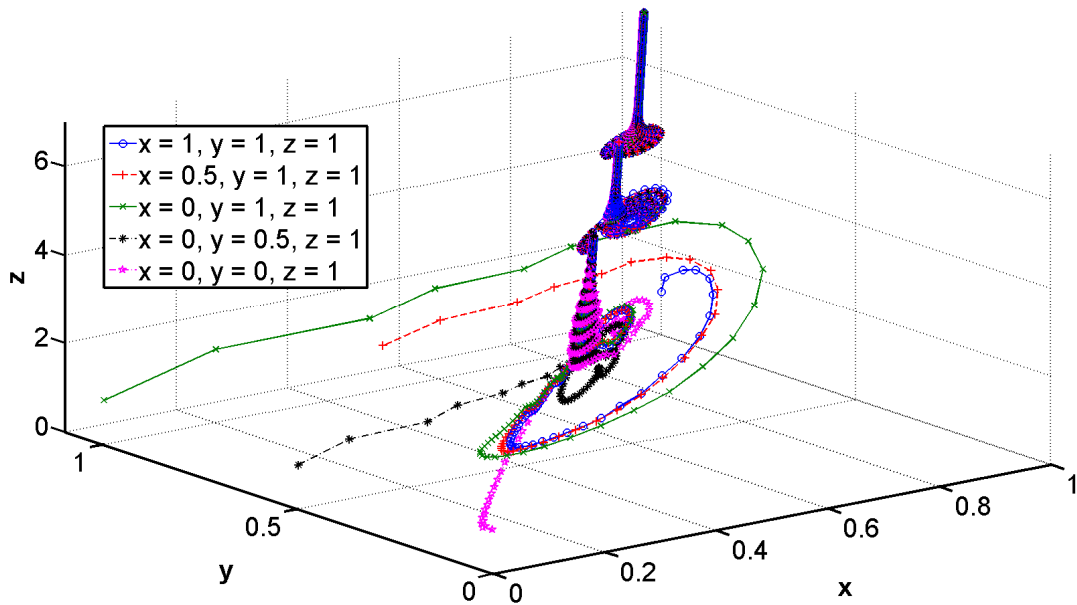

Fig. 17. Robustness of the PID controller for chaos suppression in multi-wing Lu system for different initial conditions.

In this section, the proposed GA based PID control scheme has also been tested for its robustness with variation in the initial conditions of multi-wing chaotic systems. Question may arise whether the optimum PID control scheme would be able to stabilize the system for other initial values of the state variables, since the controller is tuned with any one initial guess of the states. This is particularly important to investigate since the proposed approach does not rely on classical Lyapunov criterion based analytical stabilization which has been extensively studied for double wing attractors [19], [35], [5].

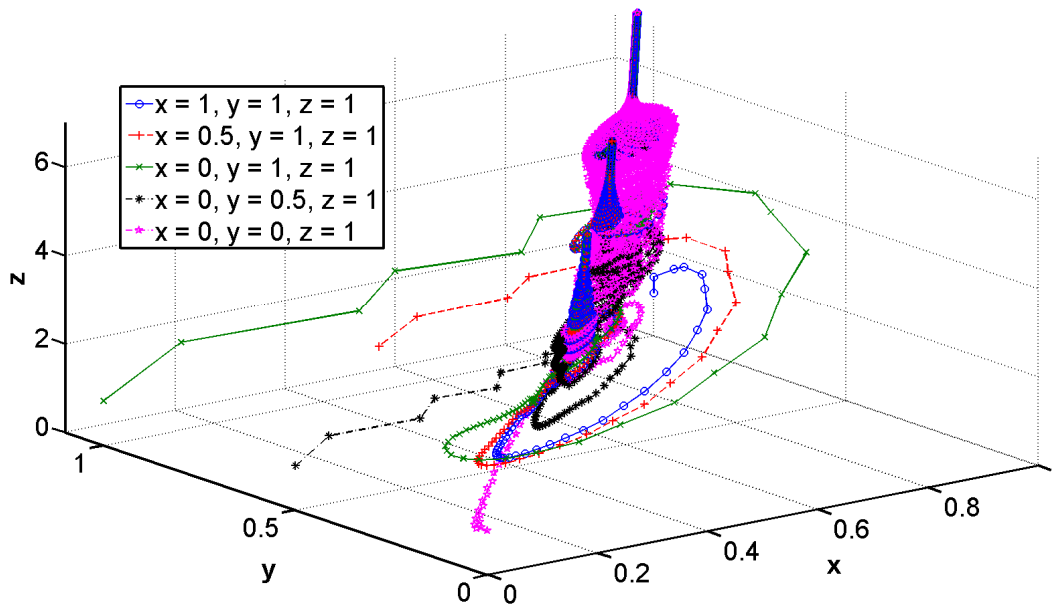

Fig. 18. Robustness of the PID controller for chaos suppression in multi-wing Chen system for different initial conditions.

Fig. 17-20 shows that in the phase portraits the chaotic oscillations get suppressed along different trajectories for the four complex nonlinear dynamical systems, even if the initial conditions for the first two states are gradually decreased from unity to zero. In spite of high sensitivity to the initial conditions for the states in all chaotic systems, the proposed PID control scheme is capable of suppressing the wandering of the states in phase space for the Lu and Chen system which are quite similar and shown in Fig. 17-18. For the Rucklidge system as shown in Fig. 19, the controlled phase space trajectories converges towards a particular direction even with variation in initial condition of the system's states. The controlled phase portraits for different initial conditions are much more complex for the Sprott-1 system as portrayed in Fig. 20 but finally converge to a stable equilibrium point, similar to the other multi-wing systems.

A comparison of Figs 17 and 18 can be made with that of Fig. 19. It can be seen that in case of the Rucklidge system in Fig. 19, the state trajectories from the different initial conditions to the final equilibrium solution are much smaller than those of the Lu or the Chen systems (in Figs. 17 and 18 respectively). This also provides an insight into the amount of controller effort and settling time required to suppress the chaotic oscillations in these systems. As discussed previously, the Lu and the Chen systems are more difficult to control and this can also be understood from the phase portraits of the trajectories in Figs. 17 and 18. In these figures the trajectories evolve through multiple small scrolls and hence take more time than that of the Rucklidge system in Fig. 19. The trajectories of the controlled systems in Fig. 20 show that they take larger excursions than the Rucklidge system but do not get trapped in small scrolls, therefore making them easier to control than the Lu and the Chen systems, but more difficult than the Rucklidge system.

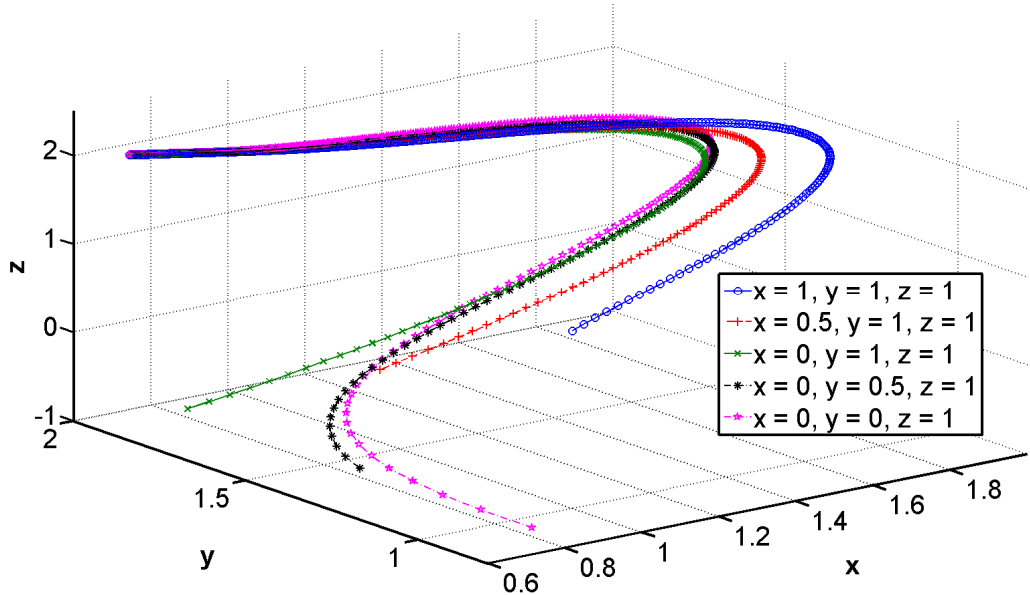

Fig. 19. Robustness of the PID controller for chaos suppression in multi-wing Rucklidge system for different initial conditions.

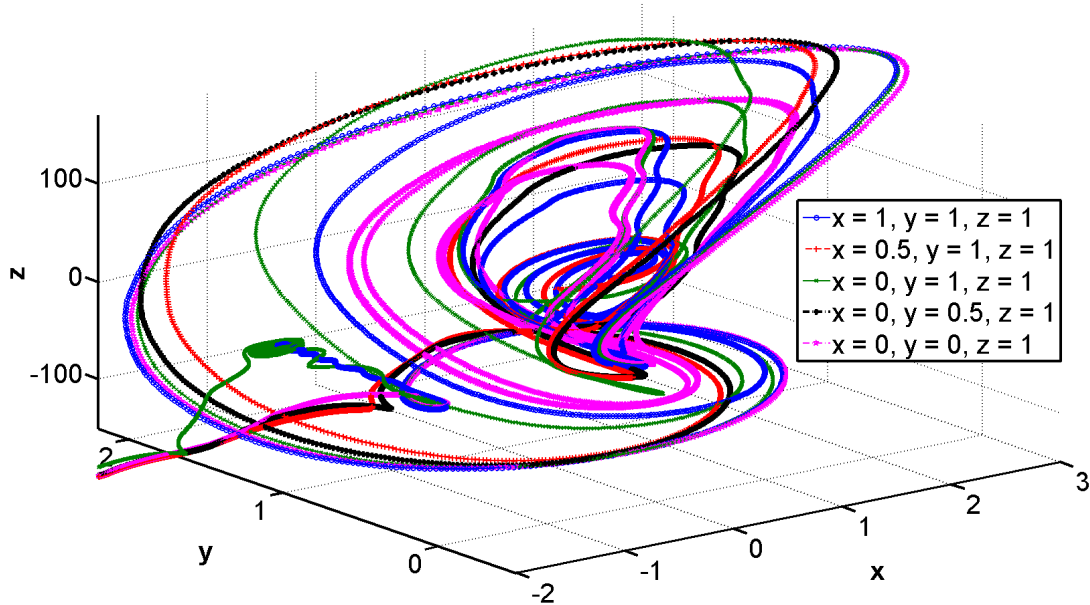

Fig. 20. Robustness of the PID controller for chaos suppression in multi-wing Sprott-1 system for different initial conditions.

## 5. Discussion:

It is well known that it is difficult to find a quadratic Lyapunov function in such complicated nonlinear dynamical systems like multi-wing chaotic attractors except few attempts like that in [30]. The present approach has been shown to work reliably well, even though analytical stabilization has not yet been addressed for such systems due to high system complexity. The proposed optimum PID control scheme has another advantage over the conventional Lyapunov function based stabilization approach i.e. the optimum tracking of the desired state variable which is difficult to achieve with the classical Lyapunov approach. With slight modification of the objective function different states or even more than one state variable can easily be enforced to track different reference inputs [11], unlike the Lyapunov based approach. The Lyapunov and sliding mode approaches enforce guaranteed stabilization but not optimum set-point tracking and also the nonlinear control law needs to be changed every time, depending on the structure of the chaotic system which restricts the designer to have a common controller structure (like a simple PID here) to control all the multi-wing systems in the Lorenz family. Thus to the best of authors' knowledge, this is the first approach to suppress chaotic oscillations in the family of multi-wing Lorenz family of chaotic systems with a simple control scheme augmented with a global optimization framework.

Another important part of the present control scheme is the course of control action to derive the control law i.e. the effect of the upper limit of the integral objective function (16). Also, the trajectory of control for different chaotic systems plays an important role here with respect to the quality of the control. Depending on the number of equilibrium points and speed of oscillations for a particular chaotic system, the control action and the associated controller gains would vary widely within the same common PID structure. Also, the upper limit of the time domain performance index needs to be chosen suitably keeping in mind the number of oscillations within a same window. For example, the multi-wing Rucklidge system has got less number of irregular oscillations with a chosen 100 sec of simulation time window thus it gets settled down very quickly. Whereas both the multi-wing Lu and multi-wing Chen system produces more number irregular oscillation (in the uncontrolled mode) within that chosen window which takes much higher time get damped in the controlled mode. The multi-wing Sprott-1 system has the slowest dynamics compared to the other three chaotic systems thus it also settles down fast.

The particular improvements of the present paper over state-of-the-art techniques are highlighted below.

- Due to the structure specific Lyapunov stabilization scheme for a particular chaotic system, the proposed PID control scheme is more suited where the control scheme does not need to be changed every time except the PID controller gains.
- Deriving an analytical control action to suppress chaotic oscillation is based on the choice of the quadratic Lyapunov candidate which guarantees stabilization but not the fastest possible stabilization. The optimum PID controller not only suppresses the oscillations but enforces optimum tracking of one or multiple state variables (with suitable choice of the objective function as in [11]).
- Most of the popular analytical control schemes for a particular chaotic system rely on manipulating all the state variables [19], [35], [5]. The present PID control scheme only senses the second state variables and manipulates it to perform the same stabilization task. This is practical from the hardware implementation point of view as less number of sensors is required. Also not all the states of the chaotic system may be measurable. In such cases, this scheme scores over the others.

- The control scheme is particularly useful even in the presence of noise in the measured variable as previously studied in [21] which is difficult tackle with using analytical methods of nonlinear state feedback law design [19], [35], [5]. The PID control scheme, due to not being dependent on the system structure and directly working on the sensed state time series instead of the system model, is capable of stabilization of such nonlinear dynamical systems [12]. This is difficult to achieve with the classical state feedback controller.

In spite of having several advantages, it is well known that stabilization of chaotic systems cannot be guaranteed by simulation due to their sensitive dependence on initial conditions. Although the simulations presented in section 4 in order to prove the robustness of PID control for multi-wing chaotic systems show promising results, still after testing with thousands of different initial conditions, the stabilization is not theoretically guaranteed unlike the Lyapunov based approach. Except the unavailability of guaranteed stabilization for all possible initial condition of the states, the proposed control scheme enjoys more flexibility of the controller design in a common template and can easily be tweaked by modifying the objective function (16) and the global optimizer used in this context.

**6. Conclusion**
A real coded Genetic Algorithm based optimum PID controller design has been proposed in this paper to suppress chaotic oscillations in a family of highly complex multi-wing Lorenz family of chaotic systems. The PID controller enforces fast tracking of the second state due to presence of the ITAE criterion in the controller design phase which also removes chaotic oscillations in other state variables. The present approach is based on a heuristic optimization framework which can be easily modified to enforce optimum tracking performance in the other state variables and also with a combination of them. Using credible numerical examples, the optimum PID control scheme has been shown to be robust enough with different initial conditions, even for such highly complex nonlinear dynamical systems known as the multi-wing Lorenz family of chaotic systems. As mentioned earlier, obtaining analytical guaranteed stabilization scheme is quite challenging for such complex dynamical systems and has been left as a scope for future research.